# Low-Temperature Sputtering and Polarity Determination of Vertically Aligned ZnO Nanocolumns

A. Hamzi[1], L. Ouardas[1], M. Saleh[1], P. Leuasoongnoen[2], T. Sonklin[3]
P. David[1], S. le Denmat[1], O. Leynaud[1], E. Mossang[1], B. Fernandez[1], S. Pojprapai[4],
D. Mornex[1], R. Songmuang[1]*

[1]Université Grenoble Alpes, CNRS, Grenoble INP, Institut Néel, 38000 Grenoble, France
[2]Synchrotron Light Research Institute, Nakhon Ratchasima 30000, Thailand
[3]Institute of Research and Development, Suranaree University of Technology, Nakhon Ratchasima 30000, Thailand
[4]School of Ceramic Engineering, Institute of Engineering, Suranaree University of Technology, Nakhon Ratchasima 30000, Thailand
*Corresponding author

Abstract

We report the low-temperature (80-100°C) growth of vertically aligned ZnO nanocolumns on Si(001) substrates by using reactive radio-frequency magnetron sputtering. High sputtering pressure combined with low substrate temperatures induce a pronounced self-shadowing effect, leading to the formation of isolated nanocolumns. In contrast, lower sputtering pressure promotes void filling in deposited films, favouring the growth of dense, low-roughness columnar films. Modification of native $SiO_x$/Si surfaces via substrate preheating prior to deposition, alters the initial nucleation stage, thereby determining dominant polarity and morphology of ZnO nanostructures. O-polar–dominated columnar films and nanocolumns exhibit higher effective piezoelectric coefficient, corresponding to their higher differential resistance and reduced dielectric loss, suggesting suppressed carrier-induced screening of piezo-charges. This low-thermal-budget, scalable sputtering approach provides an alternative route for integrating ZnO nanostructures onto thermal constrained substrates, including those used in flexible and wearable electronics.

1. Introduction

Zinc Oxide (ZnO) is a wide-bandgap semiconductor with a non-centrosymmetric wurtzite crystal structure, granting it inherent piezoelectric properties. The coupling between mechanical deformation and electrical polarization, together with its earth abundance and biocompatibility, positions ZnO as a compelling alternative to traditional piezoelectric ceramics, such as lead zirconate titanate (PZT). Owing to their advantages, ZnO is recognized as an important functional material in electronics, optoelectronics, sensing, and biomedical technologies, especially for wearable and mechanical energy-harvesting applications.

Unlike ferroelectric materials, wurtzite ZnO possesses spontaneous and piezoelectric polarizations along the c-axis, which is its preferential growth direction, without requiring high-voltage poling. Although the piezoelectric coefficient of bulk ZnO ($d_{33}$

≈12 pC/N) is approximately one order of magnitude lower than that of PZT, its dielectric permittivity (ε) is also significantly lower. As a consequence, the piezoelectric voltage constant ($g_{33}=d_{33}/\varepsilon$), which is a key figure of merit for sensing applications, becomes comparable to that of traditional piezoelectric ceramics. Moreover, the electromechanical performance of ZnO can be further enhanced by reducing its dimensionality to nanoscale. One-dimensional ZnO nanowires and nanocolumns are particularly advantageous as their high aspect ratio enables enhanced mechanical compliance and efficient strain transfer. Their sensing potential is evidenced by the remarkable sensitivity of ZnO nanogenerators to biomechanical stimuli, including finger tapping, hand clapping, and vocal vibrations[1,2,3]

A wide range of ZnO nanostructures can be synthesized using various conventional chemical and physical growth techniques[4]. Solution-based methods, including chemical bath deposition and hydrothermal growth allow low-temperature and large-area fabrication. However, these techniques often suffer from high incorporation of residual impurities that increase free carrier concentration and limit piezoelectric performance through charge screening[5,6,7,8]. Vacuum-based physical deposition techniques offer an effective route to suppress such impurities, with magnetron sputtering being particularly attractive due to its scalability and industrial compatibility. However, vertically aligned ZnO nanocolumns achieved by sputtering typically require growth temperatures exceeding 500°C[9,10,11,12], severely limiting their integration with thermally sensitive substrates such as polymers which are generally used in flexible electronics.

In this work, we present an alternative sputtering growth regime that enables the formation of vertically aligned ZnO nanocolumns on Si substrates at the growth temperatures of 80-100°C. We demonstrate that the argon flow rate, which controls sputtering pressure and gas-phase scattering, is the key parameter driving the morphological transition from dense columnar films to isolated nanocolumns under limited adatom mobility conditions. This growth transition is qualitatively consistent with the shift from the transition zone (Zone-T) to the porous zone (Zone 1) in Thornton's structural zone model[13].

Using macroscopic direct piezoelectric response measurements and valence band X-ray photoemission spectroscopy (VB-XPS), we show that the dominant polarity of these ZnO nanostructures depends on the initial nucleation conditions, which can be tuned via low-temperature pre-annealing of the native $SiO_x$ surface on Si substrates. Finally, we correlate the effective piezoelectric coefficients ($d_{33,\text{eff}}$) with the dielectric losses and differential resistances, highlighting a critical role of carrier-induced screening of generated piezo-charges which reduces the apparent piezoresponse signals. These results offer a previously under-explored growth window, which is adaptable for integrating ZnO nanostructures with flexible or wearable devices.

2. Experimental details
ZnO Growth by Reactive RF Magnetron Sputtering

ZnO nanostructures were deposited on highly boron-doped p-type Si (100) substrates with a resistivity of less than 0.01 Ω·cm via reactive radio-frequency (RF) sputtering using a ceramic ZnO target (99.99% purity). Prior to deposition, a base pressure of the sputtering chamber was below 5 × 10$^{-8}$ mbar. A 10-minute pre-sputtering step was performed on the target to remove its surface contamination. The substrates were heated to the selected growth temperature (80–100 °C unless otherwise stated) and maintained at this temperature for 1 hour to ensure thermal stabilization before deposition. The substrate temperature was monitored by using a thermocouple located at the substrate holder's backside. Depositions were carried out for 8250 seconds at an RF power of 50 W, with a fixed oxygen flow rate of 2 sccm, and a substrate-target distance of 10 cm. The argon flow rate was varied between 15 and 80 sccm.

Structural Morphological Characterization

The crystalline structure, orientation and quality were analysed by using a Bruker D8-Advance diffractometer with Bragg-Brentano Reflexion geometry, utilizing a Ge monochromator (Cu Kα1 radiation=1.54056 Å). Surface and cross-sectional morphologies were examined by field-emission scanning electron microscopy (FESEM, Zeiss Ultra55). Atomic force microscopy (AFM, Bruker DI-V) measurements were performed in tapping mode to evaluate surface roughness. Film thickness and nanocolumn height were estimated from cross-sectional FESEM images. The vertical growth rates were determined as the ratio of the total thickness from cross-sectional FESEM images to deposition time.

Piezoelectric and Electrical Characterization

The macroscopic direct piezoelectric response was evaluated using a custom-built setup based on the Berlincourt principle, described in detail elsewhere [14]. A sinusoidal mechanical excitation with an amplitude of 200 mN was applied normal to the sample surface over a frequency range of 0.16–150 Hz. A static preload of 4–6 N was used to ensure electrical and mechanical stability, without affecting the dynamic piezoelectric response signal. The piezoelectric charge response was detected by a charge amplifier (Kistler 1505A) with an operating frequency up to 30 kHz.

For electrical characterizations, circular top electrodes with a diameter of 1 mm were deposited by using e-beam evaporation of 10 nm Ni/80 nm Pt bilayer through a shadow mask. The backside contact, composed of 10 nm Ti/100 nm Au, was deposited on a highly doped Si substrate. Silver epoxy was applied to mechanically reinforce the top electrodes during dynamic loading. Following the piezoelectric response measurements, impedance measurements were performed at the same position under the same static preload using a Keysight 4980AL LCR meter in a frequency range of 20 Hz-1 MHz with an excitation voltage of 200 mV. Current-voltage (I-V) characteristics were measured using a Keithley 2636B source meter unit within a ±1 V range to determine direct current (DC) characteristics and extract differential resistance ($R_{diff}$) at a low bias of 50 mV.

X-Ray Photoemission Spectroscopy

X-ray photoemission spectroscopy (XPS) measurements were carried out at the SUT-NANOTEC-SLRI joint research facility located at the beamline 3.1 in Synchrotron Light Research Institute, Thailand. A scanning XPS Microprobe (PHI5000 Versa Probe II, ULVAC-PHI) generates X-rays from Al Kα radiation at 1486.6 eV. The beam spot size was focused to 100 μm, and photoelectrons were collected at a 45° take-off angle relative to the surface normal. The binding energy is referenced from the C1s peak (C-C/C-H) at 284.8 eV. Valence band spectra were analysed to assess the dominant polarity of ZnO nanostructures.

3. Results and discussion
Morphological Transition and Growth Regimes

The structural evolution of ZnO sputtered at a low substrate temperature of 100 °C is primarily governed by the argon flow rate, which determines the sputtering pressure. As shown in the cross-sectional FESEM and AFM images in Figure 1, a distinct morphological transition occurs between argon flow rates of 15 sccm and 30 sccm. At a low flow rate of 15 sccm, the deposited ZnO forms relatively smooth, dense columnar films with a mean roughness of around 1–2 nm. However, increasing the argon flow rate to 30 sccm and above triggers the formation of vertically separated nanocolumns with open intercolumnar boundaries.

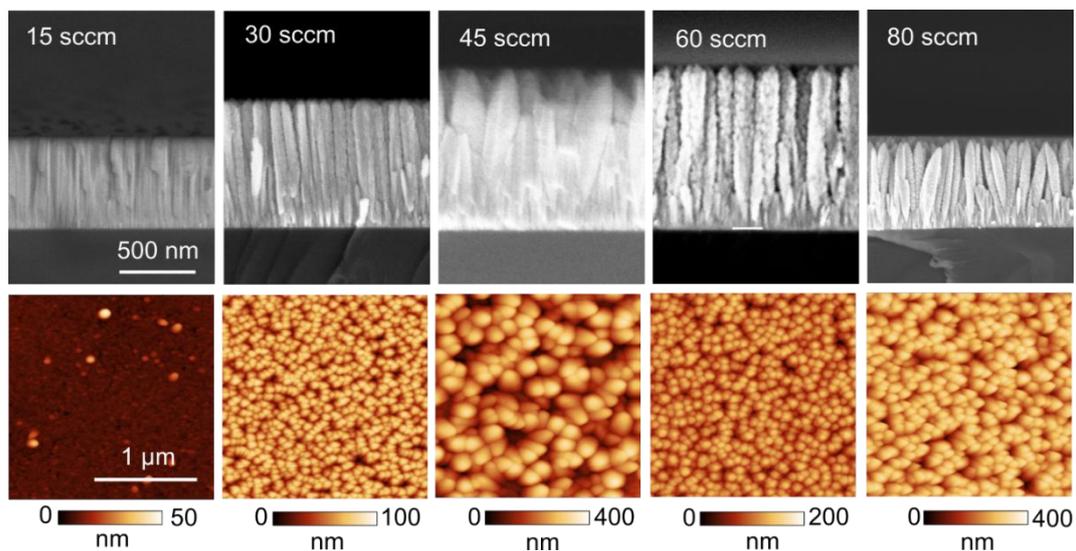

Figure 1 Morphological evolution of ZnO nanostructures sputtered on Si(100) at substrate temperature of 100 °C. Side-view FESEM (upper panel) and corresponding AFM (lower panel) images of samples grown by using Argon flow rates of 15, 30, 45, 60, and 80 sccm. The transition from dense columnar films to vertically separated nanocolumns is observed when the Argon flow rate is increased.

AFM analysis further supports this transition, revealing that the mean surface roughness of the nanocolumn ensemble is nearly one order of magnitude higher than that of the compact columnar films. The total film thickness and nanocolumn height initially increase with increasing argon flow rate and subsequently decrease when the flow rate reaches 80 sccm. Notably, nanocolumn formation is also observed at elevated substrate temperatures of 700-750 °C when the argon flow rate remains at 80 sccm, whereas only planar columnar films are obtained at 15 sccm over the wide temperature range of 80–750 °C. These observations indicate that sputtering pressure, rather than substrate temperature, is the dominant parameter governing nanocolumn formation.

The observed morphological transition is ascribed to suppressed adatom mobility combined with enhanced self-shadowing effects at high sputtering pressures. As the background pressure increases, sputtered species undergo more frequent gas-phase collisions, leading to a broader angular distribution and a larger fraction of obliquely incident atoms[15]. This non-directional flux promotes geometrical shadowing by the growing film itself, resulting in void formation. The lack of thermal energy at low substrate temperature prevents deposited atoms from diffusing across the surface to fill those structural voids, which physically separate the growing material into isolated nanocolumns. The nanocolumn formation is qualitatively consistent with the growth in Zone 1 of Thornton's structural zone model.

At lower sputtering pressures, reduced scattering yields a more directional flux normal to the substrate, facilitating lateral growth and void filling, and resulting in compact columnar films with low surface roughness. This growth regime corresponds to the transition zone (zone T) of the Thornton's model. Although this model was originally developed for metallic thin films, it can phenomenologically describe our observed scenario.

Structural Properties and Crystallographic Orientation

The normalized XRD spectrum of these samples are summarized in Figure 2(a). According to the JCPDS database, for wurtzite ZnO (No. 36-1451) and face-centred cubic Si (No. 27-1402), the $2\theta$-diffraction peak of relaxed hexagonal wurtzite ZnO(0002) is located at 34.422°, while that of the (400) plane of silicon appears at 69.132°. Thus, we attribute the diffraction peak located at around 34.5° to a ZnO (0002) plane, indicating that the c-axis of the ZnO wurtzite crystal is oriented normal to the substrate. We also observe the peaks located at 31.7° and 36.2° emerge from the samples showing separated nanocolumns, which are the diffraction peaks of the (10-10) non-polar plane and semipolar plane (10-11) of ZnO. These planes correspond to the sidewalls and the top facets of the ZnO nanocolumns, respectively.

Figures 2(b)-(c) summarize the ZnO(0002) XRD peak position and the extracted strain along the c-axis as a function of sputtering pressure and corresponding Argon flow rate, respectively. As the sputtering pressure increases, the ZnO(0002) peak systematically shifts toward 34.422° which is the position of a fully relaxed wurtzite ZnO crystal. This shift is utilized for estimating the apparent strain in the sputtered ZnO films[16].

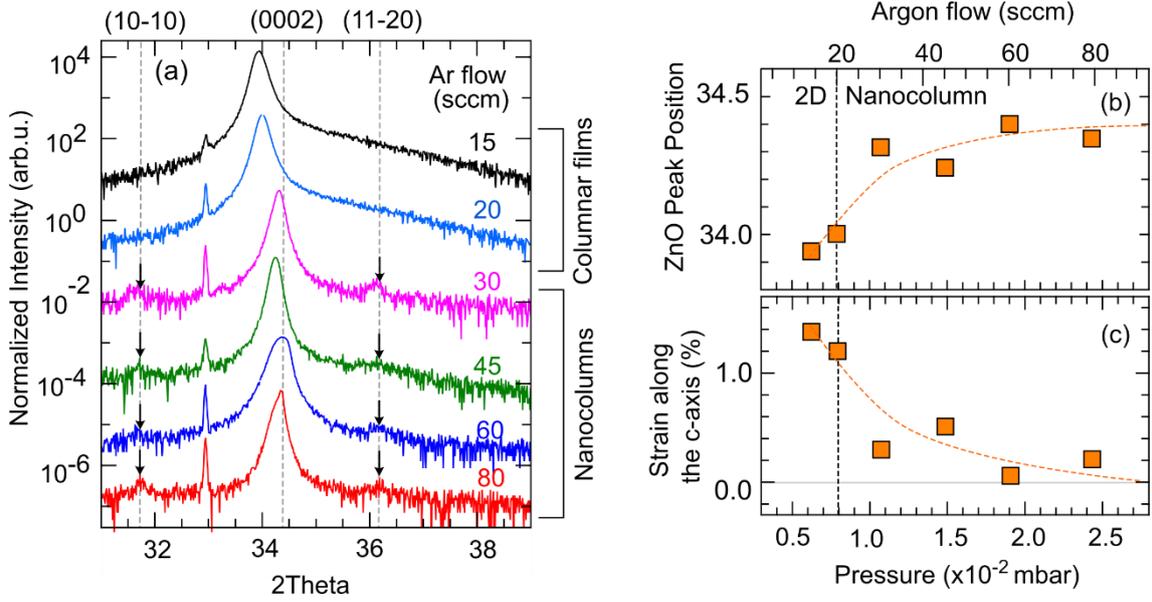

Figure 2 Structural and strain analysis of sputtered ZnO as a function of sputtering pressure and the corresponding argon flow rate. (a) XRD spectrum of the samples from Figure 1. (b) Evolution of the ZnO(0002) diffraction peak position, and (c) Extracted out-of-plane strain as a function of sputtering pressure.

The extracted data in Figure 2(c) reveals that the ZnO columnar films experience a tensile strain along the c-axis, implying an in-plane compressive strain existing in the films due to the Poisson's effect. This compressive strain is ascribed to the atomic peening mechanism during magnetron sputtering[17]. This effect is driven by energetic particles, primarily reflected Ar neutrals and sputtered atoms, which bombard the growing film surfaces and knock atoms into interstitial positions or densify the microstructures by eliminating voids. In the case of continuous films, this expansion is laterally constrained by the clamping effect of the substrates and prior layers, resulting in a biaxial compressive stress. At low substrate temperatures as the ones utilized here (100°C), the suppressed surface and bulk diffusion prevent strain relaxation via defect annealing or grain growth.

As the morphology transitions into separated nanocolumns at higher Argon flow rates, this strain systematically decreases. Unlike the continuous films, the open boundaries of the nanocolumns allow the crystal lattice to relax more effectively. Furthermore, the higher background pressure increases the frequency of gas-phase collisions, reducing the peening effect as energetic atoms lose their kinetic energies before reaching the substrates.

Polarity Determination and Nucleation Effects

The crystallographic polarity of the ZnO nanostructures was investigated using a combination of macroscopic direct piezoelectric response measurements and VB-XPS. A dynamic force was applied along the c-axis of the samples as illustrated in Figure 3(a). The ratio of the generated piezoelectric charge to the applied force provides the effective piezoelectric coefficient ($d_{33,\,\text{eff}}$) along the stressing direction.

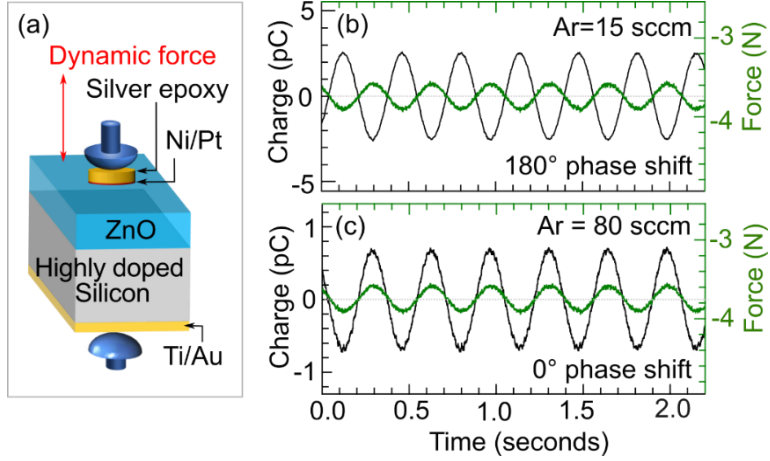

Figure 3 (a) Schematic illustration of the device structure and the measurement configuration. Piezoelectric charge response as a function of time from columnar films (b) and nanocolumns (c) grown by using Ar flow rates of 15 sccm and 80 sccm, respectively. The green curve represents a sinusoidal force excitation with an amplitude of 200 mN at 3 Hz.

Figures 3(b)-(c) display representative piezoelectric charge responses (black curves) from columnar films and nanocolumns grown at 100°C by using argon flow rates of 15 and 80 sccm, respectively. The amplitude of the sinusoidal force excitation was 220 mN with a frequency of 3 Hz (green curves). The 180° phase shift of the charge response relative to the force excitation shown in Figure 3(b) indicates the O-polar surface of the columnar film, while the 0° phase shift suggests the Zn-polar nature of the ZnO nanocolumns. These results reflect the dominant polarity under the macroscopic excitation area. Regarding the magnitude of the response from these specific samples, the columnar film exhibits $d_{33,\text{eff}}$ of around 16 pC/N which is higher than that of the nanocolumns (4 pC/N).

Figure 4(a) summarizes the phase shifts of the piezoresponse signal as a function of sputtering pressure and the corresponding argon flow rate. This figure illustrates a transition from O-polar columnar films grown at low pressure (180° phase-shift) to Zn-polar nanocolumns (0° phase-shift) grown at high pressure. In addition, VB-XPS spectra were acquired to corroborate the polarity assignment derived from the macroscopic piezoelectric measurements. As shown in Figure 4(b), the VB-XPS spectra from nanocolumns grown using the argon flow rates above 30 sccm, exhibit two distinct peaks: Peak I at ~ 5 eV associated with O 2p-derived states and Peak II at ~ 7 eV, attributed to a hybridized state of O 2p, Zn 4s, and possibly Zn 3d[18,19]. The relative intensity ratio of these two peaks serves as an indicator for ZnO polarity[20,21,22,23]. In the nanocolumns, the intensity of Peak I exceeds that of Peak II, consistent with a Zn-polar (0001) surface, whereas the columnar films lack of well-resolved peaks, suggesting an O-terminated surface. The interpretation of VB-XPS spectra agree with that of the piezoelectric phase response shown in Figure 4(a).

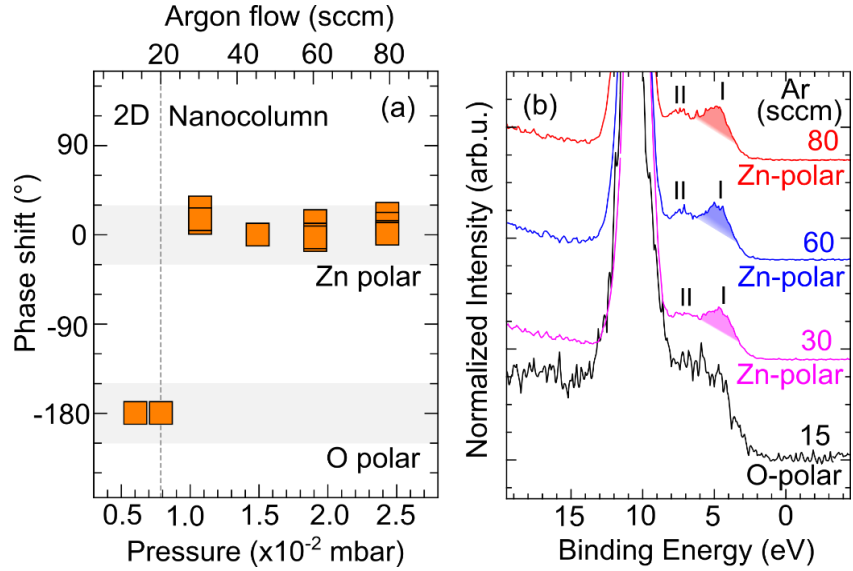

Figure 4 (a) Summary of the piezoresponse phase shifts from ZnO columnar films and nanocolumns grown at 100°C as a function of sputtering pressure and corresponding argon flow rate. (b) corresponding VB-XPS spectra with the sub-peaks indicated by filling.

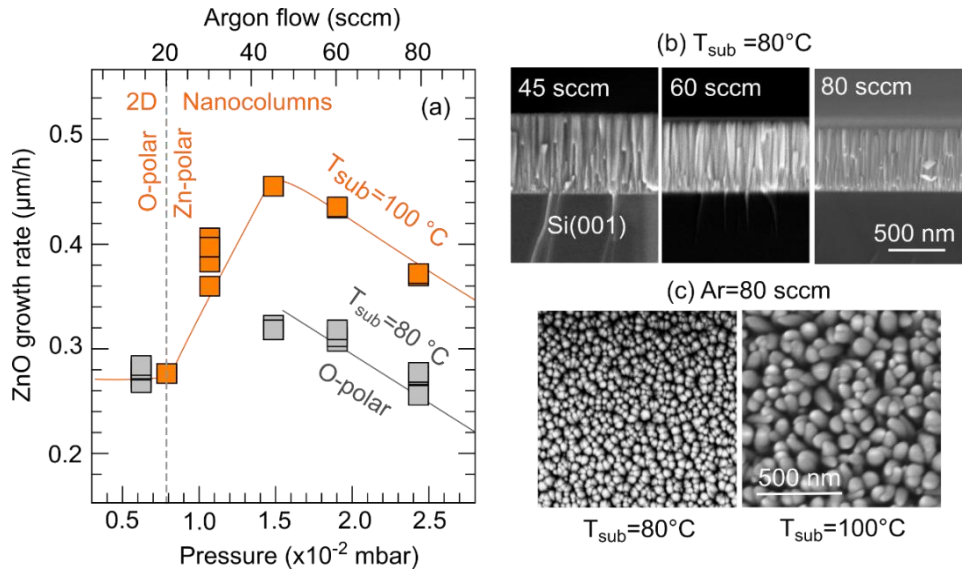

Figure 5 (a) ZnO vertical growth rate plotted as a function of sputtering pressure and the corresponding argon flow rate. (b) Side-view FESEM images of ZnO nanocolumns sputtered at a substrate temperature of 80 °C using argon flow rates of 45, 60, and 80 sccm. (c) Top-view FESEM images of ZnO nanocolumns sputtered at substrate temperatures of 80 and 100 °C with an Ar flow rate of 80 sccm.

Figure 5 summarizes the vertical growth rate of ZnO columnar films and nanocolumns as a function of sputtering pressure and the corresponding argon flow rate. Initially, the growth rate increases from 0.25 µm/hour to approximately 0.45 µm/hour at an argon flow rate of 45 sccm (see also Figure 1). This enhancement is attributed to an increased flux of sputtered Zn and O species arriving at the substrate surface, driven by a higher density of active Ar⁺ ions bombarding the target at a given RF power. However, the growth rate systematically declines, when the argon flow rate is further increased to 80 sccm. This reduction is a result of increased collision events at high pressure, which reduces both their kinetic energy and the number of active atoms reaching the growth front.

When the substrate temperature ($T_{sub}$) is reduced to 80°C, the high-pressure regime with argon flow rates of 45-80 sccm continues to promote nanocolumn formation. The resulting nanocolumn ensemble shows higher areal density, smaller diameters, and a reduced average growth rate, with a maximum value of 0.3 µm/hour [Figures 5(b)-(c)]. This increased nanocolumn density can be attributed to a larger number of nucleation sites caused by shorter adatom surface diffusion lengths at lower temperature. As a result, the incoming material flux is distributed over a greater number of growing nanocolumns, leading to a reduced axial growth rate of individual structures.

Interestingly, the nanocolumns synthesized at 80 °C exhibit an O polar surface, similar to columnar films, in contrast to the Zn-polar nanocolumns grown at 100 °C. This polarity difference further contributes to the lower growth rate of O-polar nanocolumns. As Zn-polar surfaces possess a higher density of dangling bonds than O-polar surfaces, this enhances surface reactivity and facilitates faster growth. The growth rate of Zn-polar nanocolumns is approximately 1.5 times higher than that of O-polar one, consistent with previous reports.[24,25,26].

Effect of Substrate Pre-Annealing on the Polarity of Nanocolumns

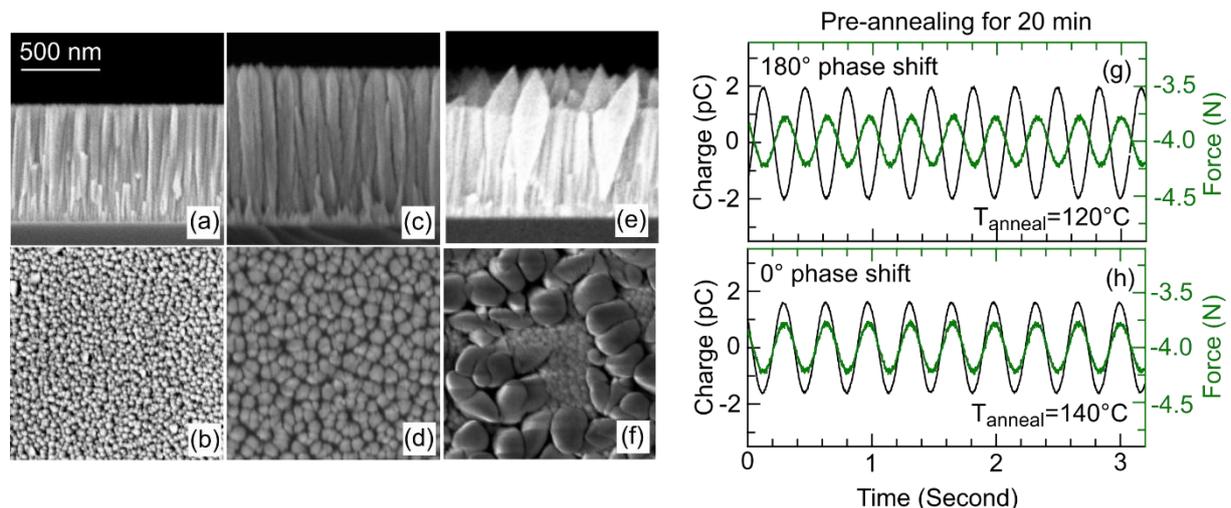

Figure 6 Side-and top-view FESEM images of ZnO nanocolumns grown on Si substrates using different pre-annealing conditions prior to ZnO sputtering i.e., (a)-(b) 120°C for 20 min, (c)-(d) 140°C for 20 min, (e)-(f) 120°C for 100 min. Piezoelectric charge responses as a function of time from nanocolumns grown by using the pre-annealing conditions of (g) 120°C for 20 min and (h) 140°C for 20 min. The green curve represents a sinusoidal force excitation with an amplitude of 200 mN at 3 Hz.

The O-polar nature of columnar films can be rationalized by total energy minimization during the initial nucleation stage. When ZnO is sputtered onto Si substrates covered with native SiO$_x$, no epitaxial relationship exists between the film and the substrate. Consequently, ZnO exhibits self-textured growth, favouring crystallographic orientations that minimize surface free energy. Initial nucleation preferentially proceeds along the <0001> direction, allowing the formation of low-surface-energy, non-polar {10–10} side facets[27,28]. Because of the lower surface energy of the O-polar surface compared to the Zn-polar one[29,30,31], it is energetically more stable during early nucleation. At low sputtering pressure with argon flow rate of 15 sccm, the efficient void-filling process, which is a characteristic of Zone T growth, freezes the O-polar orientation. This growth behaviour persists up to substrate temperatures of 300 °C.

Increasing the sputtering pressure enhances void formation and leads to nanocolumn growth. Under these conditions of low adatom mobility, O-polarity is expected to remain, as it is determined during the initial nucleation stage. However, we observe a polarity switching from O-polar to Zn-polar nanocolumns when the substrate temperature increased from 80°C to 100°C [Figure 4(a)]. Since this temperature range is insufficient to significantly alter the native SiO$_x$ morphology or enable substantial surface diffusion, we attribute this polarity inversion to changes in the substrate surface chemistry. To test this hypothesis, Si substrates were subjected to low-temperature pre-annealing step. Afterward, the substrate temperature was set at 80°C and thermally stabilized for 180 min before sputtering ZnO nanocolumns using an argon flow rate of 80 sccm.

Figures 6(a)-(f) present side-and top-view FESEM images of ZnO nanocolumns grown on substrates pre-annealed under different conditions i.e., 120°C for 20 min, 140°C for 20 min, and 120°C for 100 min. For substrates pre-annealed at temperature of 120°C for 20 min [Figures 6(a)- (b)], the nanocolumns exhibit a high areal density, small diameters, and reduced axial growth rates, closely resembling the nanocolumns grown at 80 °C without a pre-annealing step. In contrast, pre-annealing at 140°C for 20 min produces faster-growing and better-separated nanocolumns [Figures 6(c)-(d)]. When the pre-annealing at 120 °C was prolonged to 100 min, a heterogeneous morphology, featuring a mixture of compact and fast-growing nanocolumns becomes evident, indicating non-uniform nucleation conditions [Figures 6(e)-(f))].

Large-scale direct piezoresponse measurements reveal that well-separated nanocolumns exhibit Zn polar surface, whereas more compact ones are predominantly O-polar. On the other hand, a bimodal morphology reflects mixed polarity, where steeper pyramidal facets are expected to originate from faster growing Zn-polar ZnO domains[32]. This polarity inversion is attributed to variations in the chemical state of the SiO$_x$ surface which affect the initial nucleation, governed by the pre-annealing conditions.

This behaviour can be explained by the interplay between surface silanol (Si–OH) groups and physisorbed water layers[33,34], which are highly sensitive to annealing temperature and duration. After pre-annealing at 120 °C for 20 min, the native SiO$_x$ surface retains a high density of silanol groups and physically adsorbed water.

Hydrogen bonding between silanol groups and water molecules limits their direct interaction with incoming $Zn^{2+}$ species. Consequently, interfacial bonding remains weak, and ZnO growth proceeds under thermodynamic control, favouring the O-polar orientation.

Increasing the annealing temperature to 140°C for 20 min largely desorbs physisorbed water while preserving silanol groups which become chemically accessible. These Si–OH groups can form Si–O–Zn interfacial bonds with positively charged $Zn^{2+}$ions, kinetically stabilizing Zn-polar nucleation and dictating the subsequent stacking sequence. The nanocolumns grown at 100 °C without pre-annealing likely correspond to this regime.

Prolonged pre-annealing at 120 °C to 100 min induces partial dehydration, progressively exposing silanol groups in a spatially nonuniform manner. This heterogeneous interfacial chemistry results in competing nucleation pathways and mixed-polarity growth.

Notably, the O-polar ZnO columnar films grown in Zone T persists up to substrate temperatures of 300°C, despite potentially high -OH group densities at elevated temperatures. The results suggests that the O-polar surface is more stable during two-dimensional growth, ascribed to its favourable lateral growth and coalescence resulting from its lower surface energy compared to the Zn-polar surface[35].

Electrical Leakage and Piezoelectric Response

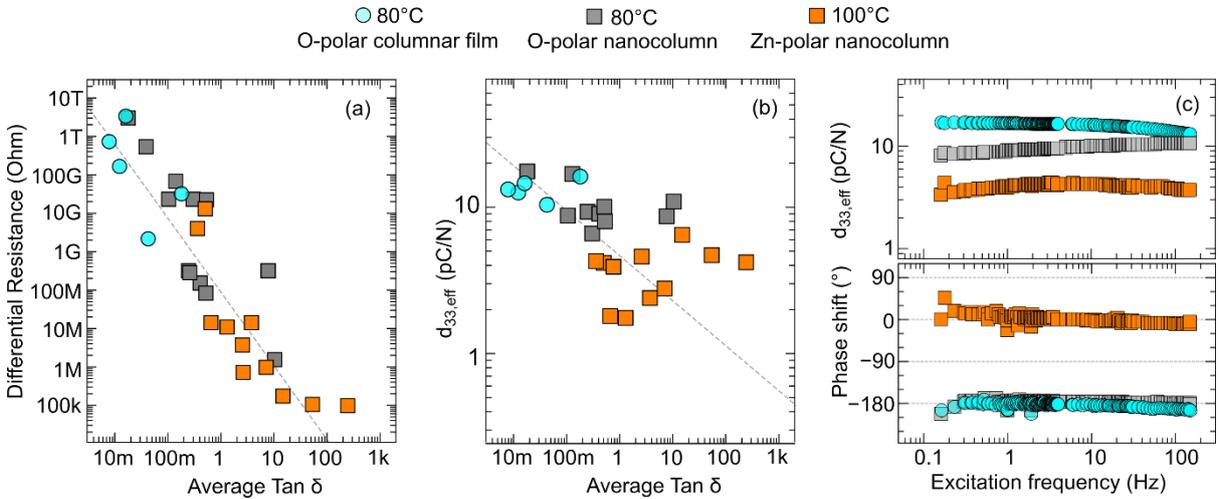

Figure 7 (a) Differential resistance extracted from I-V measurements and (b) $d_{33,eff}$ at 10 Hz plotted as a function of average tan δ over 50-150 Hz. Data taken from O-polar ZnO columnar films and O-polar nanocolumns grown at 80 °C, as well as Zn-polar nanocolumns grown at 100 °C. (c) $d_{33,eff}$ (upper panel) and phase (lower panel), as a function of excitation frequency ranging from 0.1 to 150 Hz from the selected samples measured in (a) and (b).

To evaluate the piezoelectric quality of the samples, we correlated their differential resistance extracted from I-V measurements and the $d_{33,\text{eff}}$ with the average dielectric tangent loss (tan δ) as shown in Figure 7. The I-V characteristics of the ZnO based-devices exhibit pronounced nonlinearity, consistent with charge transport across a Schottky barrier at the NiPt/ZnO interface and a p-n junction formed between n-type ZnO and $p^{++}$Si substrate. The differential resistance extracted at a low bias of 50 mV, corresponds to small-signal carrier transport relevant to low-frequency excitation. The tan δ was averaged over the 50–150 Hz, obtained from impedance spectroscopy, to match the frequency window of mechanical excitation used for piezoelectric measurements.

In ZnO, dielectric loss at low frequencies is dominated by mobile charge carriers and is thus closely related to electrical conductivity. As shown in Figure 7(a), a clear inverse correlation is observed: lower differential resistance corresponds to higher average tan δ. Differential resistances exceeding 10 GΩ (beyond the upper limit of our measurement setup), predominantly in O-polar ZnO columnar films (cyan circle symbol) and nanocolumns (grey square symbols), reflect a semi-insulating nature of low-temperature-grown ZnO caused by depletion zones and high defect densities. At higher frequencies, free carriers are unable to follow the alternating field, causing tan δ to approach an intrinsic baseline governed by lattice vibrations and defect-related dipolar relaxation[36]. Consequently, low-frequency tan δ is more reliably reflecting the screening of piezoelectric polarization in our measurement range than single-frequency measurements or resistance-only metrics.

Figure 7(b) reveals that samples with higher tan δ systematically display reduced $d_{33,\text{eff}}$ values, demonstrating that enhanced leakage promotes significant screening of piezoelectric charges. Owing to their high resistance and low tan δ, the O-polar columnar films display the highest $d_{33,\text{eff}}$ values, reaching up to 16 pC/N. In contrast, the Zn-polar ZnO nanocolumns show more scattered and generally lower $d_{33,\text{eff}}$ values, which is attributed to greater leakage associated with higher impurity incorporation on the more chemically reactive Zn-polar surface[37]. Higher free carrier concentrations in Zn-face ZnO compared to O polar-one were reported consistently in the literature[38]. The O-polar ZnO nanocolumns grown at 80 °C exhibit intermediate leakage and high, though slightly reduced $d_{33,\text{eff}}$ values compared to columnar films, likely due to additional device-level leakage pathways arising from their geometry and device fabrication. The measured $d_{33,\text{eff}}$ and the phase response remains relatively constant over the 1-150 Hz frequency range [Figure 7(c)], indicating that carrier-mediated screening occurs on timescales longer than the response time of the detection circuit. The stable piezoelectric performance in the physiological frequency range, combined with their low-temperature growth highlights their suitability for integration into wearable or biomedical sensing devices.

Conclusion

We demonstrate a low-temperature growth window (80–100 °C) for vertically aligned ZnO nanocolumns on Si(001) using reactive RF magnetron sputtering. High argon flow rates combined with low substrate temperatures promote pronounced self-shadowing, characteristic of Zone 1 in Thornton's structural zone model, resulting in the formation of separated nanocolumns. In contrast, lower sputtering pressure reduces gas-phase scattering, enabling the growth of compact columnar films with low surface roughness. Pre-annealing the Si substrates prior to the growth modifies the native $SiO_x$ surface during the initial nucleation stage, which can activate a polarity transition from O-polar to Zn-polar nanocolumns.

The $d_{33,\text{eff}}$ extracted from these ZnO columnar films and nanocolumns is primarily limited by a leakage-induced charge screening mechanism. This effect is influenced by polarity-dependent surface chemistry; for instance, the higher reactivity and impurity incorporation of Zn-polar surfaces lead to increased conductivity. Within the detection bandwidth, both the columnar films and nanocolumns exhibit a stable $d_{33,\text{eff}}$ across the low-frequency range (1-150 Hz) compatible with physiological signals. These values typically surpass those of solution-processed counterparts (2-3 pC/N). Overall, this work establishes a previously under-explored growth regime for ZnO nanocolumns at low substrate temperatures, offering a simple, scalable, and vacuum based approach compatible with thermally sensitive substrates including polymers, widely used in flexible and wearable electronics.

Acknowledgments

We acknowledge the assistance from Nanofab, Néel Institute for sample fabrications and SERAS, Néel Institute for setup development. The work was financially supported by ANR-15-IDEX-02, ANR-PRCI NanoFlex (Project No. ANR-21-CE09-0044), AAP Tremplin@Physique 2025 (Flex-Zense). We also thank the PHC SIAM 2024 mobility program (Project No. 50889RD) provided by the French Ministry of Europe and Foreign Affairs, the French Ministry of Higher Education, Research and Innovation and the Thai Ministry of Higher Education, Science, Research and Innovation. We also acknowledge BL3.1, the SUT-NANOTEC-SLRI Joint Research Facility, SLRI, Thailand for the XPS measurements.